\documentclass{new_tlp}
\pdfoutput=1
\usepackage{mathptmx}
\usepackage{wrapfig}
\usepackage{lscape}
\usepackage{rotating}
\usepackage{color}
\usepackage{url}
\usepackage{boxedminipage}
\usepackage{amsmath}
\usepackage{textcomp}
\usepackage{graphicx}
\usepackage[style=math]{k}
\definecolor{purple}{rgb}{0.5,0,0.9}

\usepackage{pgfplots}
\pgfplotsset{compat=newest}
\pgfplotsset{every tick label/.append style={font=\footnotesize}}
\pgfplotsset{testbar/.style={
    nodes near coords xbar stacked configuration/.style={},
    xbar stacked,
    width=.575\textwidth,
    axis y line*= none, axis x line*= bottom,
    xmajorgrids = true,
    xmin=0,xmax=100,
    xticklabel={\pgfmathparse{\tick}\pgfmathprintnumber{\pgfmathresult}\%},
    ytick = data, yticklabels = { Bank Model, Blocks World, BRP, Dekker, Maze, Philosophers, Rent-a-car (fm), Stock Ex., Stock Ex. (fm), Webmail app},
    tick align = outside, xtick pos = left,
    bar width=2.25mm, y=2.85mm,
    enlarge y limits={abs=0.625},
    legend style={at={(0.5,1.2)}, anchor=north,legend columns=-1,draw=none}
}}

\usepackage{adjustbox}

\def\defemb#1#2{\expandafter\def\csname #1\endcsname
                              {\relax\ifmmode #2\else\hbox{$#2$}\fi}}
\defemb{cA}{{\cal A}}
\defemb{cB}{{\cal B}}
\defemb{cC}{{\cal C}}
\defemb{cD}{{\cal D}}
\defemb{cE}{{\cal E}}
\defemb{cF}{{\cal F}}
\defemb{cG}{{\cal G}}
\defemb{cH}{{\cal H}}
\defemb{cI}{{\cal I}}
\defemb{cJ}{{\cal J}}
\defemb{cK}{{\cal K}}
\defemb{cL}{{\cal L}}
\defemb{cP}{{\cal P}}
\defemb{cR}{{\cal R}}
\defemb{cS}{{\cal S}}
\defemb{cT}{{\cal T}}
\defemb{cU}{{\cal U}}
\defemb{cV}{{\cal V}}
\defemb{cX}{{\cal X}}
\defemb{cZ}{{\cal Z}}
\newcommand{\relation}{\bullet\!\!\!\rightarrow} 

\title[Assertion-based Analysis via Slicing with {\sf ABETS}]{Assertion-based Analysis via Slicing with {\sf ABETS}\thanks{
This work has been partially supported by the EU (FEDER)
and Spanish MINECO grant TIN2015-69175-C4-1-R,
and
by Generalitat Valenciana PROMETEOII/2015/013. J.\ Sapi\~na was supported by FPI-UPV grant SP2013-0083.}\\(System Description)}

  \author[M. Alpuente et al.]
         {M. ALPUENTE, F. FRECHINA, J. SAPI\~{N}A\\
         DSIC-ELP, Universitat Polit\`ecnica de Val\`encia\\
         \and
         D. BALLIS\\
         DIMI, University of Udine\\
}
 
\pagerange{\pageref{firstpage}--\pageref{lastpage}}
\newtheorem{example}{Example}[section]
\begin{document}
\label{firstpage}
\maketitle

\begin{abstract}
We present {\sf ABETS}, an assertion-based, dynamic analyzer that helps diagnose errors in Maude programs. {\sf ABETS} uses slicing to automatically create reduced versions of both a run's execution trace and executed program, reduced versions in which any information that is not relevant to the bug currently being diagnosed is removed. In addition, {\sf ABETS} employs runtime assertion checking to automate the identification of bugs so that whenever an assertion is violated, the system automatically infers  accurate slicing criteria from the failure. We summarize the main services provided by {\sf ABETS}, which also include a novel  assertion-based facility for program repair that generates suitable program fixes when a state invariant is violated. Finally, we provide an experimental evaluation  that shows the performance and effectiveness of the system. This paper is under consideration for publication in TPLP.
\end{abstract}


\section{Introduction}
Bug diagnosis is a time-consuming and, most often, tedious   manual task that forces   developers to painstakingly examine large volumes of complex execution traces while trying to locate the actual cause of observable misbehaviors.  This paper   describes a dynamic program analyzer   called {\sf ABETS} {(``Assertion-BasEd Trace Slicer'')}, which aims to mitigate the costs of   diagnosing errors in  concurrent programs that are written in Maude. 

Maude is a language and a system that efficiently implements Rewriting Logic (RWL) \cite{Meseguer92}, which is a logic of change that seamlessly unifies a wide variety of models of concurrency. Thanks to its logical basis, Maude provides a precise mathematical model, which allows it to be used as a declarative language and as a formal verification system. Maude supports rich formal   specification, equational rewriting, and logical reasoning modulo {\it algebraic axioms} (such as associativity, commutativity, and identity), providing tools for a  number of formal techniques that include theorem proving, protocol analysis, state space exploration,  deductive verification,  model transformation, constraint solving, and model checking. The execution traces generated by Maude are complex objects     to analyze since they may contain a huge number of compound rewrite steps   that, however, omit crucial information for debugging such as the application of algebraic axioms (which is concealed within Maude's {\it equational matching} algorithm). While this maximizes efficiency and is certainly justified during the  program operation, it further complicates debugging. The dynamic analyzer {\sf ABETS}  described in this paper facilitates the   debugging of Maude programs. It does this by  drastically  simplifying the size and complexity of the analyzed programs and runs while still showing all relevant information for  debugging, which is done by a  fruitful combination of runtime assertion checking and slicing that was originally formalized in  \cite{ABFS16-jlamp}. In assertion-based slicing, {the user supplements the  Maude program to be analyzed with a set of logical assertions } that are   checked at runtime. Upon an assertion failure, an accurate set of discordant positions (called symptoms) is {{\it automatically} calculated by  {\sf ABETS}} by  comparing the {\it computed} erroneous {program state with} the   {\it expected} {pattern for the state} (as defined by the violated assertion), with the comparison being performed by using least general generalization {\it modulo} the algebraic axioms  of the operators involved \cite{AEEM14-ic}. By filtering out everything but the distilled disagreements, a so-called {\it slicing criterion} is synthesized {by {\sf ABETS}} that accurately identifies  the (position of the) faulty  information in the erroneous last state of the trace. Then, in order to locate the source of the error, a trace slicing procedure is  automatically triggered   that   propagates the anomalous information. This is done    by recursively computing the origins or  {\it antecedents} \cite{FT94}  of the observed positions while removing everything but the computed antecedents at each step. The given combination of runtime   checking and slicing yields a self-initiating, enhanced dynamic slicing technique that traverses the program execution and makes every single computation detail explicit while revealing only and all data in the trace that contribute to the   criterion observed. As a by-product of the trace slicing process,    an executable program slice   is also automatically extracted  that captures the program subset that is  concerned  with the  error. 

Assertion-based slicing  is efficiently implemented in {\sf ABETS} not just for Maude, but also for Full Maude \cite{Maude06}, which is a powerful extension of Maude that provides support for  object-oriented specification and advanced module operations. The major strength of the system is that the user needs not  identify  criteria or error symptoms in advance because the assertions (or more precisely, their runtime checks) are used to synthesize the slicing criteria. This is a significant improvement over  more traditional, hand-operated slicing in which the criteria for slicing need to be provided by the user.\\ 
 
\noindent{\it Contributions.}
The basic algorithms behind {\sf ABETS} were introduced in   \cite{ABFS16-jlamp}, where we evaluated them on a prototype  implementation of the system. This work describes the latest, fully-fledged {\sf ABETS} implementation, which improves system efficiency as well as  the generality/flexibility of the overall  technique.

\begin{itemize}
\item We explain the   functionality of ABETS  in Section \ref{sec:features}. In Section \ref{sec:core}, we describe the assertion-based trace slicing facility. In Section \ref{sec:autofix}, we outline a  new  repair technique  that automatically suggests program corrections to fix the  program faults that  are signalled by the violation of a state invariant property.  The corrected rules are  guarded by a suitable instance of the invariant so that the repaired rule is fired only if the invariant is fulfilled. In Section \ref{sec:extrafeat}, we present some novel extra analysis features that complement the {\sf ABETS core} functionality.  
\item We provide a description of those novel implementation details and optimizations that have boosted the system performance in Section  \ref{sec:optimizations}. Also, we report a new in-depth experimental evaluation of the system in Section \ref{sec:abets-performance} that assesses critical aspects such as the assertion-checking and  slicing capabilities, and the system input/output  performance, which is a usual weak spot of  tools developed in (Full) Maude.
\item  The {\sf ABETS} system is available at  \url{http://safe-tools.dsic.upv.es/abets}. It can be  downloaded  and locally installed as a stand-alone console application,  or it can be remotely used via a user-friendly web interface.  A brief discussion of related tools and concluding remarks are provided in Section  \ref{related}. 
\end{itemize}

\section{Modeling Concurrent Systems in Maude:  Our Running Example}{}\label{sec:prelim}
Concurrent systems can be formalized through Maude programs. A Maude program essentially consists of two components, $E$ and $R$, where $E$ is a canonical (membership) equational theory     that models system states as terms of an algebraic data type, and  $R$ is a set of rewrite rules that define transitions between states. Algebraic structures often involve axioms like associativity (A), commutativity (C), and/or identity (a.k.a unity) (U) of function symbols, which cannot be handled by ordinary term rewriting but instead are  handled implicitly by working with congruence classes of terms. This is why the membership equational theory $E$ is decomposed into a disjoint union  $E= \Delta \uplus Ax$,  where the set $\Delta$ consists of (conditional) equations and   membership axioms (i.e.,\ axioms that assert the type or {\it  sort} of some terms) that are implicitly oriented from left to right as rewrite rules (and operationally used as simplification rules), and  $Ax$ is a set  of algebraic axioms, implicitly expressed  as function attributes, that are only used for $Ax$-matching.

The  concurrent system evolves by rewriting   states using {\it equational rewriting}, i.e.,\ rewriting with the rewrite rules in $R$ {\it modulo} the equations and axioms in  $E$ \cite{Meseguer92}. More precisely, execution traces (i.e., system computations) correspond to rewrite sequences  $t_0\stackrel{r_0}{\longrightarrow}_{E} t_1 \stackrel{ r_1}{\longrightarrow}_{E}\ldots$, where   $t \stackrel{r}{\longrightarrow}_{E} t'$ denotes a transition (modulo $E$) from state $t$ to   $t'$  via the rewrite rule of $R$ that is uniquely  labeled with ${r}$. Assuming that the initial term $t$ is normalized (this assumption is not essential, but it will simplify the exposition), each single transition  $t \stackrel{r}{\longrightarrow}_{E} t'$  {(or \it{Maude step)}} is computed as a rewrite chain  $t \: \stackrel{r}{\longrightarrow} t''\rightarrow_{\Delta}^*(t''_{\downarrow_{\Delta}})=t'$, where $t''\rightarrow_{\Delta}^*(t''_{\downarrow_{\Delta}})$ is an equational simplification sequence  that  rewrites  $t''$ into its canonical (i.e.,\ irreducible) form $(t''_{\downarrow_{\Delta}})$ using the oriented equations in $\Delta$. Although advisedly omitted in our notation, all rewrites in the chain (either applying $r$ or any of the equations in $\Delta$) are performed {\it modulo $Ax$}. When a rewrite step from term $t$ to term $t'$ via a rule $r\in R$ must be fully characterized, we will write $t \stackrel{r,\sigma,w}{\longrightarrow} t'$  where $w$ is the position in $t$ where the rewrite occurred and $\sigma$ is the computed substitution obtained by pattern matching modulo $E$.  As usual, term positions are defined  by means of sequences of natural numbers ($\Lambda$ denotes the empty sequence, i.e., the root position). The result of {\it replacing the subterm} of $t$ at position $w$ by the term $s$ is denoted by   $t[s]_w$.

The following Maude program  will be used as a running example throughout the paper.

\begin{example}\label{ex:stock-exchange-spec} Let us introduce a (faulty) rewrite theory that specifies a  simplified\footnote{Maude's syntax is hopefully self-explanatory. Due to space limitations and for the sake of clarity, we only highlight those details of the system that are relevant to this work.  A complete Maude specification of the stock exchange model is available at the {\sf ABETS} website at  \url{http://safe-tools.dsic.upv.es/abets}.} stock exchange concurrent system, in which traders operate on stocks via limit orders, that is,  orders that  set the  upper bound (price {\it limit}) at which traders want to buy stocks. 

When the stock price equals or drops below the price limit $\tt L$, the associated order is {\it opened} and the trader buys the stocks at the current stock price. An order is automatically {\it closed} and the associated stocks are sold {either  (a)} when the stock price $\tt P$ exceeds the purchase price limit $\tt L$ plus a predetermined {\it profit target} $\tt PT$ (i.e., $\tt P-L \geq PT$), {or (b) when} $\tt L-P$ exceeds a predetermined {\it stop loss} $\tt SL$ (i.e., $\tt L-P \geq SL $).

\begin{figure}[ht]
\figrule
\begin{center}
{\footnotesize
\begin{verbatim}
eq  [prefT] : PreferredTraders = 'T2 .
cmb [premT] : tr(TID,C) : PremiumTrader if TID in PreferredTraders .
rl  [next-rnd] : R : SS | TS | OS => R + 1 : updP(R+1,reSeed(R+1),SS) | TS | OS .
crl [open-ord] : 
            R : (st(SID,P),SS) | (tr(TID,C),TS) | (ord(OID,TID,SID,L,PT,SL,closed),OS) =>
            R : (st(SID,P),SS) | (tr(TID,C - P),TS) | (ord(OID,TID,SID,L,PT,SL,open),OS) 
            if P <= L .
crl [close-ord-SL] : 
            R : (st(SID,P),SS) | (tr(TID,C),TS) | (ord(OID,TID,SID,L,PT,SL,open),OS) =>
            R : (st(SID,P),SS) |  (tr(TID,C + P),TS) | OS
            if P <= L - SL .
crl [close-ord-PT] : 
            R : (st(SID,P),SS) | (tr(TID,C),TS) | (ord(OID,TID,SID,L,PT,SL,open),OS) =>
            R : (st(SID,P),SS) | (tr(TID,C + P),TS) | OS 
            if P >= L + PT .
eq [updP] : updP(R,S,(st(SID,P),SS)) = 
            if (rndDelta(R * S) rem 2) == 0 
            then st(SID,S + rndDelta(R * S)),updP(R,S + 1,SS) 
            else st(SID,S - rndDelta(R * S)),updP(R,S + 1,SS) 
            fi .
eq [updP-owise] : updP(R,S,empty) = empty [owise] . 
\end{verbatim}
}
\end{center}

\caption{(Conditional) rewrite rules and equations modeling the stock exchange system.}
\label{fig:rules}
\figrule
\end{figure}

Within our system model, variable names are fully capitalized, while names that begin with the symbol {\tt '} are constant identifiers for traders, stocks and orders. System states have the form $ \texttt{R : SS | TS | OS}$, where  \texttt{R} is a natural number (called round) that models the market time evolution, and  \texttt{SS}, \texttt{TS}, and \texttt{OS} are sets\footnote{To specify sets of $\tt X$-typed elements, we instantiate the Maude parameterized sort \texttt{Set\{X\}}, which defines sets as associative, commutative, and idempotent lists of elements that is simply written as $\tt (e_1,\ldots,e_n)$. The empty set is denoted by the constant symbol $\tt empty$.} of stocks, traders, and orders, respectively.

Stocks are modeled as terms \texttt{st(SID,P)} with \texttt{SID} being the stock identifier and \texttt{P} being the current stock price. Traders are modeled as \texttt{tr(TID,C)},  where \texttt{TID} is the trader identifier and \texttt{C}  is the trader's available capital. We consider two classes of traders: premium traders and ordinary (or non-premium) traders. Premium traders are allowed to buy even if they run out of capital. Premium traders are identified by the conditional membership axiom {\tt premT} (see Figure \ref{fig:rules}) that simply checks whether the trader identifier belongs to the  (hard-coded) list {\tt PreferredTraders}, which in this example just contains the premium trader {\tt 'T2}.

Orders are specified by terms of the form \texttt{ord(OID,TID,SID,L,PT,SL,ST)}, which record the order identifier \texttt{OID}, the trader identifier \texttt{TID}, the stock identifier \texttt{SID}, the stock price limit \texttt{L}, the profit target \texttt{PT}, the stop loss \texttt{SL}, and the order status \texttt{ST} (which can be either \texttt{open} or \texttt{closed}). For simplicity,   an order allows {only} a single stock to be traded at a time. This is not a limitation since multiple stocks can be managed by multiple orders.
  
Basic operations of the stock exchange model (i.e., market time evolution, opening and closure of orders) are implemented via the rules and equations of Figure \ref{fig:rules}. The \texttt{open-ord} rule opens a trader order only if the  stock price $\tt P$ falls below or is equal to the order price limit $\tt L$. When the order is opened, the stock price  is subtracted from the trader's capital, thereby updating the capital. Note that, in the   set of stocks $\tt (st(SID, P), SS)$, the stock $\tt st(SID, P)$ is distinguished from all other stocks $\tt SS$ in the system.

Similarly, the \texttt{close-ord-SL} rule closes an order for the stock \texttt{SID} and removes it from the current state when the \texttt {SID} stock price \texttt{P} falls below or is equal to the $\tt L-SL$ stop loss threshold. The trader's capital then increases by the price  \texttt{P} that the trader gets for the sold stocks. The \texttt{close-ord-PT} rule is similar and closes an order when its stock price satisfies the profit target. 

Finally, the \texttt{next-rnd} rule models the time evolution by simply increasing the round number by one and then automatically updating the stock prices by means of the function  \texttt{updP}, which randomly increases or decreases the stock prices via the na\"ive pseudo-random number generator $\tt rndDelta$ that is re-seeded at the beginning of each round  with the round tick $\texttt{R+1}$.

Note that the   specification given in Figure  \ref{fig:rules} contains two sources of error. First, the function \texttt{updP} is flawed because it could generate non-positive stock prices, which are meaningless and should be disallowed. Second, the rule \texttt{open-ord} does not check if the available capital of a non-premium trader is enough to cover the order price limit. For instance, for the ordinary Trader {\tt 'T}, the following reachability goal (which can be solved in Maude via the {\tt search} command\footnote{Given a (possibly) non-ground term $\tt s$, Maude's {\tt search} command checks whether  a reduct of $t$ is an instance (modulo the program equations and axioms) of $\tt s$ and delivers the corresponding (equational) matcher as the computed solution.})

{\small
\begin{align*}
\hspace{-2.5mm}\texttt{ (1~:~st('S,8)~|~tr('T,9)~|~ord('O,'T,'S,12,4,3,closed)) =>\!\!*~R~:~SS~|~tr('T,C)~|~OS~.}
\end{align*}
}

\hspace{-3mm}computes (among other solutions) the substitution  {\tt\{R/3, SS/st('S, 12), C/-3, OS/ord('O, 'T, 'S, 12, 4, 3, open)\}}  that witnesses the existence of an execution trace that starts from the specified initial state and ends in a final state with a faulty, negative capital {\tt C=-3}.
\end{example}

\section{Assertion-based Program Analysis and Repair with {\sf ABETS}}\label{sec:features} {\sf ABETS} implements an automated trace slicing technique based on  \cite{ABFR14-scp} that facilitates the analysis of Maude programs by  drastically reducing the size and complexity of entangled, textually-large execution traces. The technique first uncovers data dependences within the execution trace $\cT$ w.r.t. a slicing criterion  (i.e., a set of selected symbols in the last state of $\cT$) and then produces  a trace slice $\cT^\bullet$ of $\cT$ in which  pointless information that is detected to be irrelevant w.r.t.\ the chosen criterion (i.e.,\ symbols in $\cT$ that are not origins or {\it antecedents} of the observed symbols) is replaced with the special variable symbol $\bullet$. 

Unlike the original trace slicing methodology of  \cite{ABFR14-scp} where the slicing criterion must be {\it manually} determined in advance by the user, {\sf ABETS} encompasses a runtime assertion-checking mechanism (which is built on top of the slicing engine) that was originally formalized in  \cite{ABFS16-jlamp} and preserves the program semantics. This mechanism allows the slicing criteria to be {\it automatically} inferred from falsified assertions, thereby offering more automatic support to the analysis of erroneous programs and traces.

The slicing algorithm employs {\it unification} to implement the origin-tracking procedure that properly tracks back  the   data dependences   along the trace, and the {\it generalization} (i.e.,\ anti-unification) algorithm  {\it modulo} axioms of \cite{AEEM14-ic}   to automatically identify semantic disagreements of the program behavior w.r.t.\ the assertions \cite{ABFS16-jlamp}.

{\sf ABETS} is also provided with  an automatic program repair facility, which is described in Section \ref{sec:autofix}, that suggests  fixes to potentially buggy rewrite rules whenever it detects a faulty system state of a trace $\cT$ that does not satisfy a system assertion $S\{\varphi\}$. Roughly speaking, the technique transforms the rewrite rule that is responsible for the system assertion failure (i.e.,\ the last applied rule in $\cT$ that causes (a piece of) the transformed state to match the state pattern $S$). This fix is done by adding a constrained instance of the logic formula $\varphi$ into the conditional part of the rule, which is computed by using Maude's built-in $E$-unification \cite{DEEM+16}. 

\subsection{Assertion-based Slicing in {\sf ABETS}}\label{sec:core} {\sf ABETS} supports  two types of assertions: system assertions and functional assertions.\\

\noindent \textbf{i)} {\it System} assertions: Their general syntax is $S\{\varphi\}$, where  $S$ is a term (called {\it state template}), and $\varphi$ is a logic formula in conjunctive normal form $\varphi_1\wedge\ldots\wedge\varphi_n$.

A system assertion $ S \{\varphi\} $  defines a state invariant that must be satisfied by all system states that match (modulo the equational theory $E$) the state template $S$.  When a system state $s$ does not satisfy a system assertion $ S \{\varphi\} $, the position $p$ in $s$, which is called {\it bug} position, precisely indicates the subterm of $s$ that matches $S$ and is responsible for the assertion violation. 

\begin{example}\label{ex:system}
The following system assertion specifies that the capital of ordinary traders must be non-negative in every system state of the trace:
{\small
\begin{verbatim}
R:Nat : SS:Set{Stock} | tr(TID:TraderID,C:Int),TS:Set{Trader} | OS:Set{Order} 
                            {ordinary(tr(TID:TraderID,C:Int)) implies C:Int >= 0}
\end{verbatim}
}
\noindent where the user-specified predicate {\tt ordinary(T)} simply checks whether   $\tt T$ is a non-premium trader in   the Maude program of Example~\ref{ex:stock-exchange-spec}.
\end{example}

\noindent \textbf{ii)} {\it Functional} assertions: Their general form is $\mathit{I} \: \{\varphi_\mathit{in}\} \rightarrow \mathit{O} \: \{\varphi_\mathit{out}\}$ where $I,O$ are terms, and $\varphi_\mathit{in},\varphi_\mathit{out}$ are   logic formulas. Intuitively, functional assertions specify  pre- and post-conditions over  the equational simplification $t\rightarrow_{\Delta}^*(t_{\downarrow_{\Delta}})$ that heads the rewriting $t \stackrel{r}{\longrightarrow}_{E} t'$ of any term $t$ in the system trace by providing: (i) an input template $I$ that $t$ can match  and a pre-condition $\varphi_\mathit{in}$ that $t$ can meet; (ii) an output template $O$ that the canonical form $(t_{\downarrow_{\Delta}})$ of $t$ has to match and a post-condition $\varphi_\mathit{out}$ that   $(t_{\downarrow_{\Delta}})$   has to meet (whenever the input term $t$ matching $I$ meets $\varphi_\mathit{in}$). 

\begin{example}\label{ex:functional}
Consider again the Maude program of Example~\ref{ex:stock-exchange-spec}. The functional assertion 
\begin{align*}
\texttt{\small  updP(R:Nat,S:Nat,(st(SID:StockID, P:Int),SS:Set\{Stock\})) \{ P:Int > 0 \} } \\
 \texttt{\small  -> (st(SID:StockID, P':Int),SS':Set\{Stock\}) \{ P':Int > 0 \} }
\end{align*}
specifies that stock market fluctuations modeled by function \texttt{updP}  should generate  positive stock prices provided that the input stock prices are also positive.

\end{example}

The satisfiability of the provided assertions can be checked  in two different modalities, either as a {\it synchronous} (and trace-storing) procedure that incrementally executes, checks, and potentially slices execution traces at runtime, or as an {\it asynchronous}  (off-line) procedure that processes a previously computed execution trace  against the set of provided assertions. In {\sf ABETS}, system traces can be easily generated by providing both an initial and a final reachable state. As for  equational simplification traces, they can be generated by simply providing the initial term, which is then simplified to its irreducible form.  
  
Synchronous as well as asynchronous assertion checking is implemented via  equational rewriting that  automatically reduces all matched assertions to Boolean truth values. 

\begin{figure}[h]
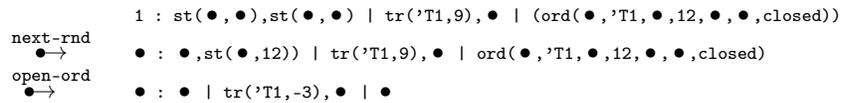

\figrule
\begin{center}
$
\begin{array}{ll}
&{\texttt{\scriptsize 1~:~} \texttt{\scriptsize st(}\bullet\texttt{\scriptsize ,}\bullet\texttt{\scriptsize ),st(}\bullet\texttt{\scriptsize ,}\bullet\texttt{\scriptsize )}\texttt{\scriptsize ~|~tr('T1,9),} \bullet \texttt{\scriptsize~|~(}}{\small \texttt{\scriptsize ord(}\bullet\texttt{\scriptsize ,'T1,}\bullet\texttt{\scriptsize ,12,}\bullet\texttt{\scriptsize ,}\bullet\texttt{\scriptsize ,closed))}}\\

{\small\stackrel{\texttt{\scriptsize next-rnd}}{\relation}~} &
{\bullet \texttt{\scriptsize~:~}\bullet\texttt{\scriptsize ,st(}\bullet\texttt{\scriptsize,12))~|~tr('T1,9),}\bullet\texttt{\scriptsize~|~ord(}\bullet\texttt{\scriptsize,'T1,}\bullet\texttt{\scriptsize ,12,}\bullet\texttt{\scriptsize ,}\bullet\texttt{\scriptsize ,closed) }} \\

{\small\stackrel{\texttt{\scriptsize open-ord}}{\hspace{-3mm}\relation}~~} &
{\bullet\texttt{\scriptsize~:~}\bullet\texttt{\scriptsize~|~tr('T1,-3),}\bullet\texttt{\scriptsize~|~}\bullet} 
\end{array}
$
\end{center}
\caption{Trace slice for automatically synthesized  criterion \texttt{tr('T1,-3)}.}
\label{fig:traceslice}
\figrule
\end{figure}

\begin{example}\label{ex:trace-slicing}

Consider the Maude program of Example~\ref{ex:stock-exchange-spec} and the execution trace $\cT =  s_0 \stackrel{\texttt{next-rnd}}{\longrightarrow} s_1 \stackrel{\texttt{open-ord}}{\longrightarrow} s_2$ that starts in the initial state
$$\begin{array}{ll}
	 {\footnotesize s_0=} \hspace{-2mm}& {\texttt{\footnotesize 1~:~(st('S1,23),~st('S2,8))~|~(tr('T1,9),~tr('T2,20))~|~ord('O1,'T1,'S2,12,4,3,closed)}}
\end{array}
$$ 
\noindent and ends in the state
$$\begin{array}{ll}
	 {\footnotesize s_2=}\hspace{-2mm}& {\texttt{\footnotesize 2~:~(st('S1,4),~st('S2,12))~|~(tr('T1,-3),~tr('T2,20))~|~ord('O1,'T1,'S2,12,4,3,open)}}
\end{array}
$$ 
The negative capital of the ordinary trader \texttt{'T1} in the state $s_2$ is demonstrably wrong  by the violation of the system assertion of Example \ref{ex:system}. Hence, {\sf ABETS} automatically computes the slicing criterion $\texttt{tr('T1,-3)}$ that pinpoints this faulty information and produces the trace slice $\cT^\bullet$ of Figure~\ref{fig:traceslice}, which represents a partial view of the system evolution  that focuses on {\tt T1}'s trading actions and exposes the erroneous behaviour of the {\tt open-ord} rule to user inspection.

\end{example}
{\sf ABETS} also  provides a handy way to automatically synthesize refined slicing criteria by means of special variables (whose name  begins with {\tt $\sharp$}) that can be used in the assertions to indicate pieces of the matched term that the user does not want to observe along the generated trace slice. For instance, if we replace {\tt TID:TraderID} with {\tt $\sharp $TID:TraderID} in the system assertion of Example  \ref{ex:system}, we compute the refined criterion $\texttt{tr(}\bullet\texttt{,-3)}$ for the trace $\cT$ of Example \ref{ex:trace-slicing}.

\begin{figure}[h!]
\centering
\includegraphics[width=\linewidth]{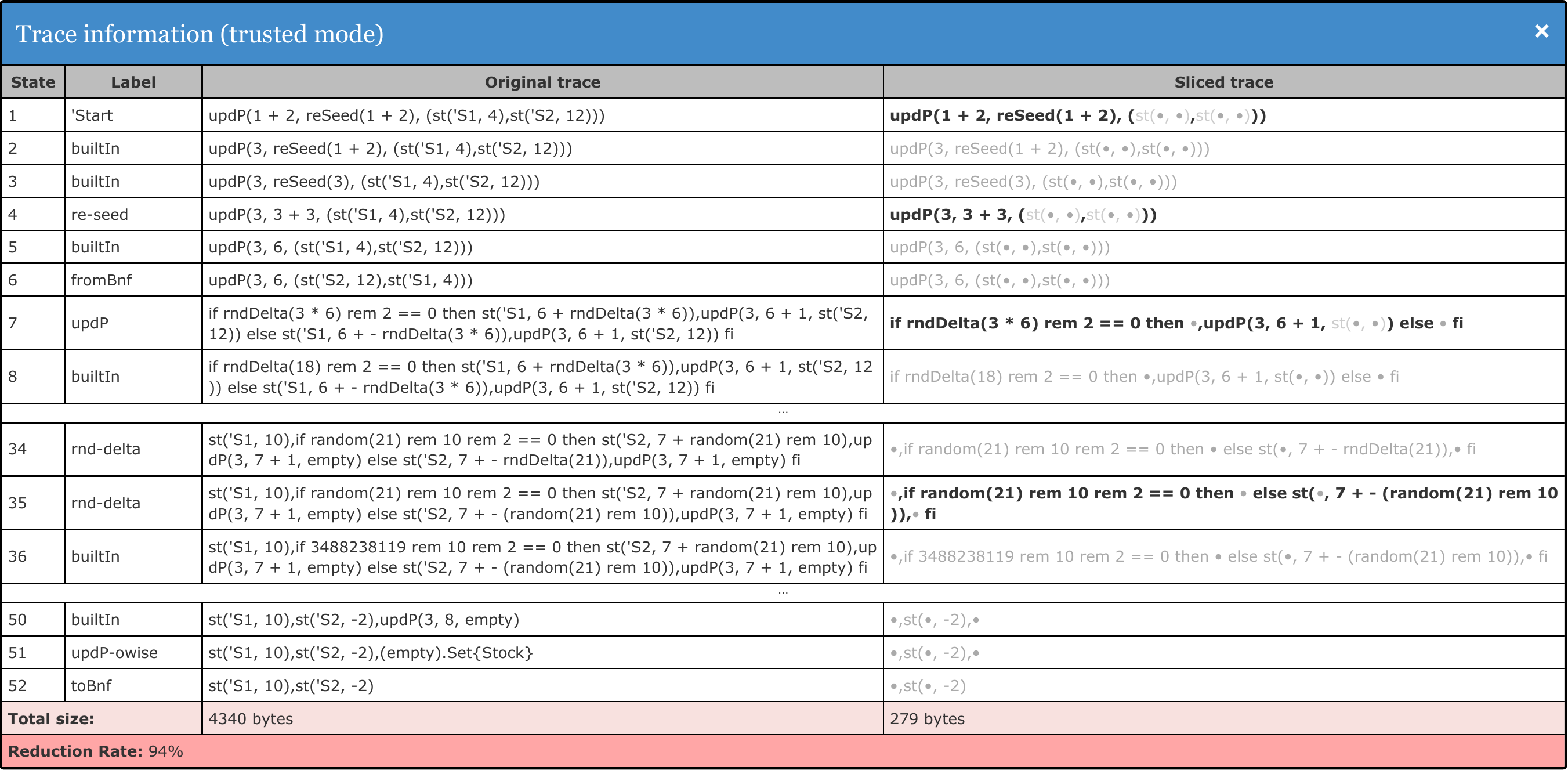}

\caption{Extended view of the computed trace slice after refuting the functional assertion of Example \ref{ex:functional} (trusted mode).}\label{fig:trustedview}

\end{figure}

\subsection{Automatic Repair of Program Rules in ABETS}\label{sec:autofix} Given an equational theory $E=\Delta\cup Ax$ and two terms $t_1$ and $t_2$, an $E$-unifier for $t_1$ and $t_2$ is a  substitution $\sigma$ such that $t_1\sigma=_E t_2\sigma$.  In Maude, $E$-unifiers are not represented as a single substitution, but as a pair of substitutions $(\sigma_1,\sigma_2)$, one for left unificands and the other for right unificands (i.e.,\ $t_1\sigma_1=_E t_2\sigma_2$). Also, Maude's $E$-unification algorithm may generate new (fresh) unification variables, denoted by \texttt{\%n}, with $\tt n$ being a natural number.  The set of all such variables contained in a given term $t$ is denoted by $\mathit{UnifVar}(t)$. Let us see an example. 
\begin{example}
Consider a simple Maude program whose signature consists of  two unary operators, \texttt{m} and \texttt{c}, and one commutative, binary operator \texttt{f}. The program includes a single equation 
\texttt{m(X)} $=$ \texttt{c(X)}. 
Then, $\sigma=(\sigma_1,\sigma_2)=({\tt \{X/\texttt{\%1}\},\{Z/\texttt{\%1}\}})$ is an $E$-unifier for the terms  $\tt t_1=\texttt{f(m(X),0)}$ and $t_2=\tt \texttt{f(0,c(Z))}$. The new, unification variable \texttt{\%1} is used to establish that \texttt{X} and \texttt{Z} represent the same value, and it is the only  common variable shared by   $t_1\sigma_1$ and $t_2\sigma_2$.
\end{example}
 
Our repair technique is based on a two-phase algorithm that takes as input: (i) the last Maude step $t \stackrel{{\tt r},\sigma,w}{\longrightarrow} t' {\rightarrow}_\Delta^{*}  t'_{\downarrow \Delta}$ of the execution trace that violates  $S\{\varphi\}$, with ${\tt r}$ being $\lambda$ \verb+=>+ $\rho$ \verb+if C+,  (ii) the violated system assertion $S\{\varphi\}$, and iii) the bug position $p$ in the last trace state $t'_{\downarrow \Delta}$.\\

\noindent{\bf Phase 1 [Semantic unification of the failing assertion and rule].} First we  $E$-unify the terms  $t'[\rho]_w$ (that is, a more general version of $t'=t[\rho\sigma]_w$ that does not apply the substitution $\sigma$ to the reduced term) and   $t'_{\downarrow \Delta}[S]_p$ (that is, a more general version of $t'_{\downarrow \Delta}$ where the subterm at the bug position $p$ is replaced by the assertion pattern $S$ itself) in order to relate the variables in the right-hand side $\rho$ of \texttt{r} with the variables that appear in the state template $S$. Since there may be several $E$-unifiers, we just select an $E$-unifier $(\sigma_\rho,\sigma_S)$ such that the bindings in $\sigma_\rho$ do not clash with the bindings in the computed substitution $\sigma$. This is done by performing a standard consistency check through the parallel composition of $\sigma_\rho$ and $\sigma$, which computes the  most general unifier (mgu) of the set of all the equations $x = t$ that represent a binding $x/t$ in either $\sigma_\rho$ or $\sigma$. If such an mgu exists, $\sigma_\rho$ is consistent w.r.t.\ $\sigma$, and the corresponding $E$-unifier ($\sigma_\rho,\sigma_S)$  is selected. 

As a side note, observe that we cannot simply $E$-unify $\rho$ with $S$ because the state template $S$ could include operators that are not in $\rho$ but in   $t'$, and, hence, the two terms could be not $E$-unifiable and lead to no repair. This is the reason why we need to $E$-unify $\rho$ and $S$ within their corresponding state contexts, that is,  $t'[\rho]_w$ and  $t'_{\downarrow \Delta}[S]_p$.  

\begin{example}
Consider a Maude program that contains the rewrite rule \verb+rl [r] f(X) => g(X)+ and no equations, together with the execution trace $\texttt{a \& f(0)} \stackrel{\texttt{r}}\longrightarrow \texttt{a \& g(0)}$ and  the system assertion \texttt{(a \& g(Z))\,\{Z>0\}},  which is violated in the state \texttt{a \& g(0)}. Observe that there is no $E$-unifier between the right-hand side \texttt{g(X)} of \texttt{r} and the state template \texttt{a \& g(Z)}, whereas the pair (\{\texttt{X/\%1}\},\{\texttt{Z/\%1}\}) is an $E$-unifier for the terms \texttt{a \& g(X)} and \texttt{a \& g(Z)}, which include \texttt{g(X)} and \texttt{a \& g(Z)} in their corresponding state context. More importantly, the bindings in the computed $E$-unifier enforce  \texttt{X} and \texttt{Z} to bind the very same value. This suggests to us that we can achieve a repair by forcing the rewrite rule argument \texttt{X} to  inherit the constraints on \texttt{Z}.
\end{example}

\noindent{\bf Phase 2 [Strengthening the rule condition].} Given the computed $E$-unifier $(\sigma_\rho,\sigma_S)$,  first we split  $\sigma_\rho$  into two sets $\sigma_{rule}$ and $\sigma_{new}$ such that  $\sigma_{rule}=\{x/t\in \sigma_\rho\mid x\in Var(\rho) \wedge\mathit{UnifVar}(t)=\emptyset \}$, and $\sigma_{new}=\sigma_\rho\setminus\sigma_{rule}$. Note that  $\sigma_{new}$ contains all those  $\sigma_{\rho}$ bindings that introduce new unification variables, while the bindings of $\sigma_{rule}$ only use the original variables of $\rho$. Then, we replace the faulty rule \verb+r+ with the following corrected rule whose condition is strengthened by adding a constrained version (that is built by using $\sigma_S$ and $\sigma_{rule}$) of the violated logic formula $\varphi$:
 $$  \verb+crl [rfix] : + \lambda\sigma_{new} \verb+ => + \rho\sigma_{new} \verb+ if C+\sigma_{new} \verb+ /\+ \Big(\big(\bigwedge_{x/t\in \sigma_{rule}} x\:\verb+==+\:t\big) \verb+ implies + \varphi\sigma_S\Big) .$$ 
 The corrected rule \texttt{rfix} is produced by instantiating the original rule \texttt{r} with the substitution $\sigma_{new}$ that introduces in \texttt{rfix} the fresh variables generated during the unification process of Phase 1 and by adding the instance  $\varphi\sigma_S$ of the falsified logical formula $\varphi$. The variables of such an instance are constrained via a logical implication whose premise is the conjunction of all the bindings $x/t$ in $\sigma_{rule}$ interpreted as Boolean expressions $x\:\verb+==+\:t$\footnote{ A binding $x/t$ in $\sigma_{rule}$ can always be interpreted as an executable, Boolean expression $x\:\texttt{==}\:t$, since all the variables included in $x/t$ appear in the rewrite rule as well and thus take concrete values when the rule is applied.} . In the case when $\sigma_{rule}$ is empty, the logical implication corresponds to  (\texttt{true implies} $\varphi\sigma_S$), and thus simply reduces to the term $\varphi\sigma_S$.

\begin{example}\label{ex:repair}
Consider a Maude program that includes the following rewrite rule \texttt{r} and equation \texttt{e}
\begin{verbatim}
    crl [r] : f(X,Y) => c(2,g(X,Y)) if X =/= Y .  
    eq  [e] : g(X,Y) = m(X,Y) .
\end{verbatim}
and assume that the operator \texttt{m} is declared commutative. Let  us consider the system assertion \texttt{c(2,m(Z,5)) \{even(Z)\}}, where \texttt{even(Z)} checks if \texttt{Z} is an even natural number.

The execution trace $\tt f(5,3)\stackrel{\tt r,\sigma}{\longrightarrow} c(2,g(5,3))\stackrel{\tt e}{\longrightarrow}c(2,m(5,3))$, with computed substitution $\sigma=\{\texttt{X/5},\texttt{Y/3}\}$, is erroneous since  the formula  \texttt{even(Z)} does not hold for the binding $\texttt{Z/3}$ that is computed by matching modulo   {\it commutativity}  the state \texttt{c(2,m(5,3))} in the   assertion state template \texttt{c(2,m(Z,5))}.
 
The repair proceeds by first performing Phase 1, which computes two $E$-unifiers of the terms \texttt{c(2,g(X,Y))} and \texttt{c(2,m(Z,5))}, namely,
\begin{align*}
(\sigma_{\rho_1},\sigma_{S_1})=(\{\texttt{X/\%1},\texttt{Y/5}\},\{\texttt{Z/\%1}\})\qquad
(\sigma_{\rho_2},\sigma_{S_2})=(\{\texttt{X/5},\texttt{Y/\%1}\},\{\texttt{Z/\%1}\})
\end{align*}
Now, observe that the $E$-unifier $(\sigma_{\rho_1},\sigma_{S_1})$ is discarded since $\sigma_{\rho_1}$ is not consistent w.r.t.\ $\sigma$. Actually, there is no  mgu of $\sigma_{\rho_1}$ and $\sigma$ because of the clash between the bindings $\texttt{Y/5}\in \sigma_{\rho_1}$ and $\texttt{Y/3}\in \sigma$.  The $E$-unifier $(\sigma_{\rho_2},\sigma_{S_2})$ is consistent w.r.t.\ $\sigma$ and thus is used to infer the repair in Phase 2 of the algorithm. 

Phase 2 generates the partition $\sigma_{\rho_2}=\sigma_{rule}\cup\sigma_{new}=\{\texttt{X/5}\}\cup\{\texttt{Y/\%1}\}$ and uses it together with 
$\sigma_{S_2}$ to yield the following corrected version of the rule \texttt{r}:
{\small
\begin{verbatim}
crl [rfix] : f(X,%1) => c(2,g(X,%1)) if (X =/= %1 /\ (X == 5 implies even(%1)) .
\end{verbatim}
} 
\end{example}   

Note that the  generated condition of a repaired rule {\tt rfix} might not be satisfiable, which makes {\tt rfix} not applicable. This is not  bad since the non-applicability of the corrected rule prevents the system from reaching the faulty state signaled by the assertion violation. This therefore has the inherent effect of reducing the number of erroneous runs in the system, which is of primary importance in the repair of critical systems as first advocated  by \cite{LB12}.

\subsection{New Additional Analysis Features}\label{sec:extrafeat}

The system functionality of   {\sf ABETS} has been extended by introducing the following, new additional features.\\

\noindent\textbf{Trusted/Untrusted modes.} 
ABETS encompasses two slicing modes: trusted and untrusted. In trusted mode, Maude built-in operators are considered to be trusted (i.e.,\  not to have bugs) and are therefore ignored in the trace slice (See Figure \ref{fig:trustedview}), which further reduces its size. In untrusted mode, all relevant operators are traced. The trusted mode is set to true by default and can be switched to untrusted mode by    choosing the {\tt Trace Information} option in the main menu  and then clicking the  {\tt Trusted/Untrusted mode} button. To help the user compare the original, extended trace and the   trace slice when they are shown side-by-side (e.g., in the table view), trusted reduction steps (as well as  duplicate states {\it modulo axioms}) are not omitted but  are depicted  in light gray.

\begin{wrapfigure}{r}{0.5\textwidth}
  \begin{center}
   \includegraphics[width=0.5\textwidth]{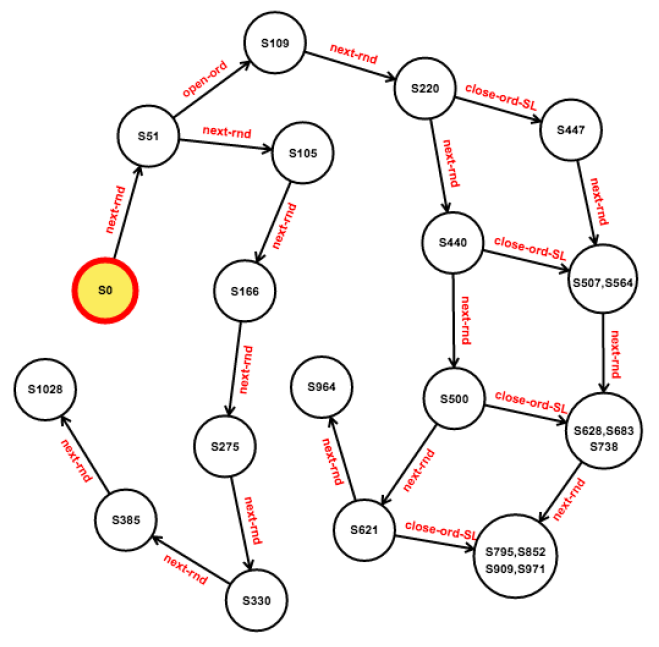}
      \caption{Computation graph generated from initial state $s_0$ of Example \ref{ex:trace-slicing} (partial view).}\label{img:graph}
  \end{center}
\end{wrapfigure} 

\noindent\textbf{Computation graph exploration.}   
To help identify traces of interest for asynchronous checking, {\sf ABETS} supports two  different representations of the computation space for a given initial term: the (standard) tree representation  that is provided by default and   a novel graph   representation of the state space that can improve user's understanding of the program behavior (see Figure~\ref{img:graph}). It is possible to switch between the two representations by left-clicking on any node of the tree or graph. In the case when the user left-clicks on a   node in the graph, the topmost leftmost node in the tree that is associated with the considered graph node is highlighted.\\

\begin{wrapfigure}{l}{0.65\textwidth}
\includegraphics[width={0.65\textwidth}]{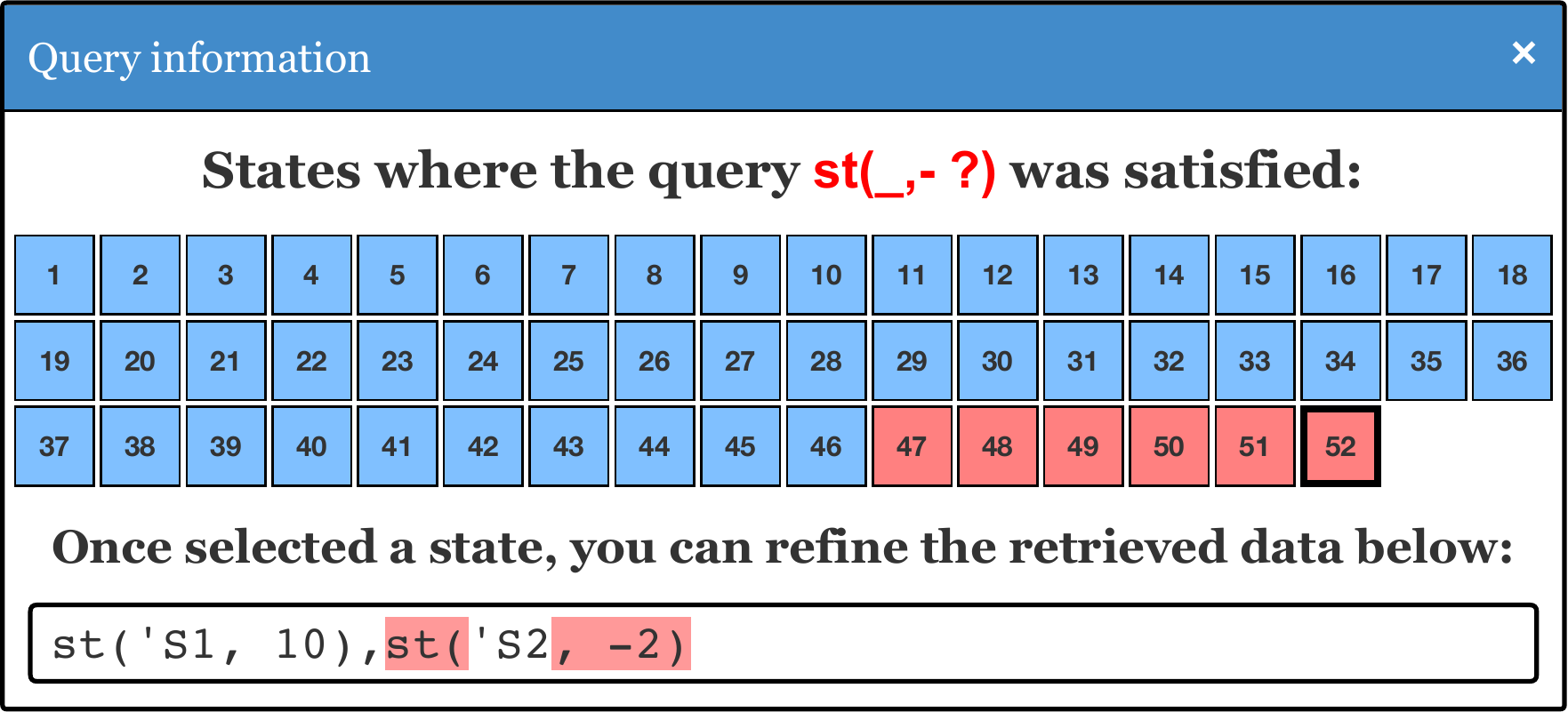}
\caption{Result of the trace query \texttt{st(\_, - ?).}}\label{img:query}
\end{wrapfigure}

\noindent\textbf{Trace querying and manipulation.} This feature  allows information of interest to be searched in huge execution traces  by undertaking a query that specifies a template for the search (see Figure \ref{img:query}).  This query is a filtering pattern with wildcards that define  irrelevant terms  by  means of the underscore character ({\tt \underline{\phantom{m}}}) and relevant  terms by means of the question mark character ({\tt ?}). In addition, traces and trace slices can  be manipulated using their meta-level representation to be exported to  other Maude tools. The  meta-representation of terms can be visually displayed, which is particularly useful for  the analysis of  object-oriented computations   where some object attributes can only be unambiguously visualized in the meta-level (desugared)   states.

Several extra features, described in \cite{ABFS16-jlamp}, are: (i) an {\it incremental trace slicing} capability that allows the computed trace slices to be further simplified    by automatically applying backward as well as forward trace slicing  w.r.t.\ user-provided slicing criteria refinements \cite{ABFS15-jsc}; (ii) a {\it program slicing} feature for delivering program fragments that include all and only the rules/equations responsible for the detected error. 
 
A starting guide that contains a typical analysis session with {\sf ABETS} can be found at \linebreak \url{http://safe-tools.dsic.upv.es/abets/quickstart.pdf}.

\section{Implementation Details and Optimizations}\label{sec:optimizations} 
The architecture of {\sf ABETS} consists of the following: (i) a Maude-based slicer and constraint-checker core that can run  at both Maude and Full Maude levels interchangeably; (ii) a scalable, high-performance NoSQL database powered by MongoDB that endows the tool with \textit{memoization} capabilities in order to improve the response time for complex and recurrent executions; (iii) a RESTful Web service written in Java that is executed by means of the Jersey JAX-RS API; and (iv) an intuitive user interface that is based on AJAX technology and written in HTML5 canvas and Javascript. {\sf ABETS}  contains about 3500 lines of Maude code,  1000 lines of C++ code, 1000 lines of Java code, and 3000 lines of Javascript code. The system   has been (re-)implemented by primarily focusing on its performance, including improvements for both the analysis and for the input and output operations.\\

\textbf{Analysis optimizations.}
One of the many features of {\sf ABETS} is its ability to manipulate all the relevant information regarding the application of equations, algebraic axioms, and built-in operators at the meta-level, which is a feature that is not supported by Maude. We implemented this extension in a new developer version of the Maude system called  Mau-Dev (available at \url{http://safe-tools.dsic.upv.es/maudev}) without affecting the   efficiency of the  latest Maude 2.7 release.  Also, to boost the system performance, the functions that are more frequently used in {\sf ABETS} have been reimplemented in C++    as new,  highly efficient, built-in  Mau-Dev (meta-level) operations that are available at  Mau-Dev's website. \\

\textbf{I/O optimizations.} Maude's efficient parser allows very large initial calls  to be efficiently parsed in just a few milliseconds. In contrast, Full Maude's parser is entirely developed in Maude itself; hence,   its efficiency can be seriously penalized   when dealing with mixfix operator definitions  due to extensive backtracking. As a result,   {\sf ABETS}  initial   calls that contain large and complex execution traces as arguments typically took  some minutes to be loaded into our previous system \cite{ABFS16-jlamp}. We have overcome  this drawback by dynamically creating a   devoted module  that defines   unique \textit{placeholder} terms that are subsequently reduced to the actual  arguments of the initial (Full Maude) call.  For example, to encode a Full-Maude, source-level representation of the  state $s_{2}$ of Example \ref{ex:trace-slicing}, {\sf ABETS} defines the 0-ary operator {\tt aState}:

\begin{verbatim}
  op aState : -> String .
  eq aState = "1 : (st('S1,23), st('S2,8)) | (tr('T1,10), tr('T2,20)) |
   (ord('O1,'T2,'S1,10,6,4,open), ord('O2,'T1,'S2,12,4,3,close))" . 
\end{verbatim}

\noindent This   greatly reduces the size of the initial Full-Maude call since it only contains the {\tt aState} placeholder but not the actual state data. These data are later brought back by applying the {\tt aState} equation. A similar encoding is  used for  user-defined assertions and execution traces that are to be  asynchronously checked.

The added module is loaded prior to starting the Full Maude's {\it execution loop} \cite{Maude06}. Thus, by taking advantage of the ability of Full Maude to access previously loaded Maude modules, the entire call can be parsed directly in Maude, except for  its top-most operator. 

The output of {\sf ABETS} executions typically consists of a Maude term of sort {\tt String}, represented in JSON (JavaScript Object Notation) format, that collects all the computed information  (e.g., the source-level and meta-level representation of the original trace and the trace slice, the associated program slice {that can be computed as described in \cite{ABFS16-jlamp}}, and  transition information between subsequent trace states). This output string is later processed by the {\sf ABETS} front-end to offer a more friendly, visual representation. Since efficient  output handling is crucial not to penalize the overall  performance  of the system, (meta) string conversion has also been  implemented in C++.

Some experiments that highlight the efficiency gain  of the optimized system  w.r.t. \cite{ABFS16-jlamp} are shown in Section \ref{sec:abets-performance}.
 
\section{Experimental evaluation}\label{sec:abets-performance}  
To evaluate the performance of the {\sf ABETS} system, we introduced defects in several Maude programs endowed with assertions  and we used the system to detect assertion violations. We benchmarked {\sf ABETS} on the following collection of Maude programs, which are all available and fully described within the {\sf ABETS} Web platform: {\it Bank model}, a conditional  Maude specification that models  a distributed banking system; {\it Blocks World}, a Maude encoding of the classical AI planning problem that consists of setting  one or more vertical stacks of blocks on a table using a robotic arm; {\it BRP}, a Maude implementation of the Bounded Retransmission Protocol; {\it Dekker},  a Maude specification of   Dekker's  mutual exclusion algorithm; {\it Maze}, the nondeterministic Maude specification of a maze game where multiple players walk, jump, or collide while trying to reach a given exit point; {\it Philosophers}, a  Maude specification of the classical Dijkstra concurrency  example; {\it Rent-a-car (fm)}, a {\it Full Maude} program that models the logic of a  distributed, object-oriented, online car-rental store; {\it Stock Exchange}, the running example of this article; {\it Stock Exchange (fm)}, a {\it Full Maude}, object-oriented version of the {\it Stock Exchange} example; {\it Webmail}, a Maude specification of a rich webmail application that provides  typical  email management, system administration capabilities,  login/logout functionality,  etc. {\sf ABETS}  automatically identifies theories that do not require Full Maude capabilities so that the highest possible analysis performance is achieved  without incurring   unnecessary costs.

\begin{table}[h!]
\vspace{-0.2cm}
  \caption{Synchronous assertion-checking performance analysis}
  \label{fig:performance}
  \begin{adjustbox}{width=1\textwidth}
  {\footnotesize
    \begin{tabular}{l|cccc|c|ccc|ccc}
    \hline\hline
    \it Program  	&$\it T_{Ex}$		&$\it T_{ExChk}$ 	&\it \#Chk	&\it OV 	&\it OV$_{\scriptsize  jlamp }$ &$\it T_{\mathit{synth}}$  & $\it T_{\mathit{fix}}$		&$T_{\mathit{I/O}}$		&\it Size $\cT_{\varepsilon}$  &\it Size $\cT_{\varepsilon}^\bullet$  &\it \%Red. \\\hline
    Bank Model      		&17		&101	&2004	&4.94 	&5.76	&2    &2 &10		&9.536      &1.236      &87\% \\
    Blocks World    		&19		&37		&509	&0.95 	&2.16	&1    &1 &2		&0.279      &0.046		&84\% \\
    BRP             		&5		&23		&1002	&3.6 	&4.6	&1    &2 &9  	&0.792      &0.269     	&67\% \\
    Dekker          		&40		&98		&1002	&1.45 	&2.5	&2    &14 &55 	&8.268      &0.286      &97\% \\
    Maze            		&128	&409	&7437	&2.2	&3		&1    &3 &13  	&2.747     	&0.423      &85\% \\
    Philosophers   			&12		&36		&811 	&2		&2.92	&1    &3 &47    	&5.244    	&1.990   	&62\% \\
    Rent-a-car (fm)			&178	&263	&1503	&0.48	&0.52	&5    &9 &247    &5.507     	&0.115      &98\% \\
    Stock Ex.       		&36		&103	&1503	&1.86 	&2.58	&3    &12 &263   	&46.423    	&4.153      &91\% \\
    Stock Ex. (fm)  		&726	&1310	&2004	&0.8	&1.72	&5    & 43 &4688   &195.397   	&20.862    	&89\% \\
    Webmail app     		&138	&271	&1002	&0.96 	&1.99	&9    &20 &541 	&133.460   	&7.823      &94\% \\
    \hline\hline
    \end{tabular}
    }
    \end{adjustbox}
\vspace{-0.3cm}
\end{table}

In our experiments, we evaluate both the effectiveness and the performance of {\sf ABETS}  by (synchronously) checking each program against an assertional specification that contains at least one failing assertion. This way,  an erroneous execution trace $\cT_{\varepsilon}$  is delivered and subsequently simplified into a trace slice $\cT_{\varepsilon}^\bullet$ w.r.t.  slicing criteria that are automatically inferred. The experiments were conducted on a PC with 3.3GHz Intel Xeon E5-1660 CPU with 64GB R{A}M.

Obviously, the  slowdown of the entire checking process  depends on   the number of assertions that are contained  in the specification and particularly on the degree of instantiation of their associated patterns. Patterns that are too general can result in a   large number of (often) unprofitable  evaluations of the logic formulas involved since the number of possible matchings (modulo axioms) with the system's states   can    grow  quickly. The slowdown can also be affected by the complexity of the predicates involved in the  functional and system assertions to be checked.

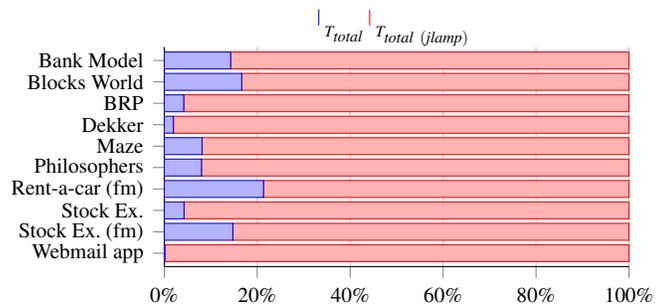
\begin{wrapfigure}{r}{0.625\textwidth}
\vspace{-0.3cm}
\begin{tikzpicture}
   \begin{axis}[testbar] 
    \addplot coordinates{
       (14.29,9) 
       (16.67,8) 
       (4.19,7) 
       (1.97,6) 
       (8.13,5) 
       (8.01,4) 
       (21.37,3) 
       (4.25,2) 
       (14.77,1) 
       (0.14,0)

    }; 
     \addplot coordinates{
       (85.71,9) 
       (83.33,8) 
       (95.81,7) 
       (98.03,6) 
       (91.87,5) 
       (91.99,4) 
       (78.63,3) 
       (95.75,2) 
       (85.23,1) 
       (99.86,0)
    };
\legend{$\scriptscriptstyle \it T_{total}$,   $\scriptscriptstyle \it T_{total~(jlamp)}$}
\end{axis}
\end{tikzpicture}
\caption{Total speedup with respect to {\scriptsize (Alpuente et al. 2016)}.}\label{fig:io}
\end{wrapfigure} 

Table  \ref{fig:performance} summarizes our results. The  $\it T_{Ex}$ and $\it T_{ExChk}$ columns measure the execution times (in ms) with and without assertion checking for traces that apply 500 rewrite rules (which expands to 8292 rewrites ---i.e.,\ rule, equation, built-in operator, and axiom applications--- on average). {\it \#Chk} represents the total number of assertion checks performed when assertion checking was enabled. {\it OV}  is the overhead, i.e.,\ the ratio $=\it (T_{ExChk}-T_{{Ex}})/T_{Ex}$ which indicates the relative slowdown due to assertion-checking. The  results obtained are quite satisfactory and comparable with similar logic assertion checking  frameworks such as \cite{MLH09}. The average overhead is 1.92, which is 69\% of the average value (2.78) of the overhead of \cite{ABFS16-jlamp} that are shown in column {\it OV$_{jlamp}$}  for the very same benchmark programs.

The figures in the  $T_{synth}$ and  $T_{fix}$ columns respectively measure the times for synthesizing the slicing criterion and for inferring the  repairs (in ms). Our experiments show very small synthesis times for  the slicing criteria that grow linearly with the size of the  erroneous state. This is particularly evident in the case of {Webmail App}, whose states are quite large (about 2.5Kb, which is 20 times the size of the Stock Ex. states). The time  for inferring the repairs is also   a small portion of the total execution time. 

The  trace slices that are  automatically delivered by {\sf ABETS} are evalutated by comparing the size of  the detected erroneous execution trace $\cT_{\varepsilon}$ (in kilobytes); the size of  the sliced execution trace $\cT_{\varepsilon}^\bullet$ (in kilobytes); and the  derived {\it reduction} rate achieved  ({\it \%Red.}), which ranges from 98\% to 62\% with an average reduction rate of 85\%. With regard to the time required to perform the slicing, our implementation is quite time efficient despite the complex analyses and reasoning modulo axioms performed underneath; the elapsed times are small even for very complex traces and also  scale linearly. For example, running the slicer for a typical  50Kb faulty trace delivered by the analyzer

Finally, the generation, parsing, and input/output of traces (and trace slices) have been greatly improved in  the current version of {\sf ABETS}. The input/output (I/O) times are shown  in column $T_{\mathit{I/O}}$ of   Table  \ref{fig:performance} (in ms) for  I/O  data sizes that range from 15~Kb (in the case of the {\tt Blocks} program) to 7~Mb (in the case of the {\tt Stock Ex.(fm)} program). This gives an average I/O cost of  0.6~s, whereas  in our previous tool  the I/O operations took minutes.

The total speedups that we achieved w.r.t.\ our previous  implementation (including checking, slicing, and I/O costs) are represented in Figure \ref{fig:io},   with an  average speedup of 9.66 with respect to  \cite{ABFS16-jlamp}.
 
\section{Conclusion and Related Work}\label{related} {\sf ABETS} combines run-time assertion checking and automated (program and execution trace) transformations for improving the debugging of programs that are written in (Full) Maude.

Assertions have been considered in (constraint) logic programming, functional programming, and functional-logic programming (see \cite{MLH09,Chitil11,AH12} and references therein). However, we are not aware of  any assertion-based, dynamic slicing system that is comparable to {\sf ABETS} for  either declarative or imperative languages. Actually, none of the correctness tools   in the related literature integrate trace slicing and assertion-based reasoning to automatically identify,  simplify,   inspect,   and repair faulty code and runs.

A  detailed discussion of the  literature related to this work can be found in \cite{ABFS16-jlamp,ABFR14-scp}. Here, we  focus on  assertion-checking tools supporting  logical reasoning modulo axioms, which are the closest to our work.

In \cite{DRMA14}, the validator tool {\sf mOdCL} is described that checks  OCL assertions on UML models encoded as Maude prototypes.

If a constraint is violated, the execution is aborted and an error is reported that signals the state and the constraint involved. In contrast to {\sf ABETS},  {\sf mOdCL} does not  simplify (either manually or automatically) the execution trace that reaches the erroneous state or the program itself  in any way. 
  
The (rewriting logic) semantic framework \K\ \cite{Rosu15} supports assertion-based analysis and runtime verification   based on  Reachability Logic (RL), a particular class of first-order formulas  with equality ${\cal P} \Rightarrow {\cal P}'$, where ${\cal P}$ (and ${\cal P}'$) consists  of a (Boolean) term $b$ and a constraint $\varphi$ over the logical variables  of $b$  (i.e.,\ $b \wedge \varphi$). These formulas  ${\cal P}$ specify those concrete configurations  that match the   algebraic structure of $b$ and satisfy the constraint $\varphi$. They are used to express (and reason about) static state properties, similarly to our system assertions $S \{\varphi\}$. As for our  functional assertions $\mathit{I} \: \{\varphi_\mathit{in}\} \rightarrow \mathit{O} \: \{\varphi_\mathit{out}\}$, they are   quantifier-free and evaluated on equational simplifications, while RL formulas assert more general properties on system computations and are used for deductive and algorithmic verification.

{A different semantic approach for automatic program  repair  that is based on abstract interpretation  can be found in \cite{LB12}, which applies to {\sf .Net} languages.}

\bibliographystyle{acmtrans}
\bibliography{biblio}

\begin{thebibliography}{}

\bibitem[\protect\citeauthoryear{Alpuente, Ballis, Frechina, and
  Romero}{Alpuente et~al\mbox{.}}{2014}]{ABFR14-scp}
{\sc Alpuente, M.}, {\sc Ballis, D.}, {\sc Frechina, F.}, {\sc and} {\sc
  Romero, D.} 2014.
\newblock {Using Conditional Trace Slicing for improving Maude Programs}.
\newblock {\em Science of Computer Programming\/}~{\em 80, Part B}, 385 -- 415.

\bibitem[\protect\citeauthoryear{Alpuente, Ballis, Frechina, and
  Sapi{\~n}a}{Alpuente et~al\mbox{.}}{2015}]{ABFS15-jsc}
{\sc Alpuente, M.}, {\sc Ballis, D.}, {\sc Frechina, F.}, {\sc and} {\sc
  Sapi{\~n}a, J.} 2015.
\newblock {Exploring Conditional Rewriting Logic Computations}.
\newblock {\em {Journal of Symbolic Computation}\/}~{\em 69}, 3--39.

\bibitem[\protect\citeauthoryear{Alpuente, Ballis, Frechina, and
  Sapi{\~n}a}{Alpuente et~al\mbox{.}}{2016}]{ABFS16-jlamp}
{\sc Alpuente, M.}, {\sc Ballis, D.}, {\sc Frechina, F.}, {\sc and} {\sc
  Sapi{\~n}a, J.} 2016.
\newblock {Debugging Maude Programs via Runtime Assertion Checking and Trace
  Slicing}.
\newblock {\em {Journal of Logical and Algebraic Methods in Programming}\/}.
\newblock To appear.

\bibitem[\protect\citeauthoryear{Alpuente, Escobar, Espert, and
  Meseguer}{Alpuente et~al\mbox{.}}{2014}]{AEEM14-ic}
{\sc Alpuente, M.}, {\sc Escobar, S.}, {\sc Espert, J.}, {\sc and} {\sc
  Meseguer, J.} 2014.
\newblock {A Modular Order-Sorted Equational Generalization Algorithm}.
\newblock {\em {Information and Computation}\/}~{\em 235}, 98--136.

\bibitem[\protect\citeauthoryear{Antoy and Hanus}{Antoy and Hanus}{2012}]{AH12}
{\sc Antoy, S.} {\sc and} {\sc Hanus, M.} 2012.
\newblock {Contracts and Specifications for Functional Logic Programming}.
\newblock In {\em {Proc. of the 14th Int'l Symposium on Practical Aspects of
  Declarative Languages (PADL 2012)}}. Lecture Notes in Computer Science, vol.
  7149. {Springer-Verlag}, 33--47.

\bibitem[\protect\citeauthoryear{Chitil}{Chitil}{2011}]{Chitil11}
{\sc Chitil, O.} 2011.
\newblock {A Semantics for Lazy Assertions}.
\newblock In {\em Proc. of the 20th ACM SIGPLAN Workshop on Partial Evaluation
  and Program Manipulation (PEPM 2011)}. {Association for Computing Machinery},
  141--150.

\bibitem[\protect\citeauthoryear{Clavel, Dur{\'a}n, Eker, Lincoln,
  Mart{\'i}-Oliet, Meseguer, and Talcott}{Clavel et~al\mbox{.}}{2007}]{Maude06}
{\sc Clavel, M.}, {\sc Dur{\'a}n, F.}, {\sc Eker, S.}, {\sc Lincoln, P.}, {\sc
  Mart{\'i}-Oliet, N.}, {\sc Meseguer, J.}, {\sc and} {\sc Talcott, C.} 2007.
\newblock {\em {All About Maude: A High-Performance Logical Framework}}.
\newblock {Springer-Verlag}.

\bibitem[\protect\citeauthoryear{Dur{\'a}n, Eker, Escobar, Mart{\'i}-Oliet,
  Meseguer, and Talcott}{Dur{\'a}n et~al\mbox{.}}{2016}]{DEEM+16}
{\sc Dur{\'a}n, F.}, {\sc Eker, S.}, {\sc Escobar, S.}, {\sc Mart{\'i}-Oliet,
  N.}, {\sc Meseguer, J.}, {\sc and} {\sc Talcott, C.} 2016.
\newblock {Built-in Variant Generation and Unification, and their Applications
  in Maude 2.7}.
\newblock In {\em {Proc. of the 8th International Joint Conference on Automated
  Reasoning (IJCAR 2016)}}. Lecture Notes in Computer Science, vol. 9706.
  {Springer-Verlag}, 183--192.

\bibitem[\protect\citeauthoryear{Dur{\'{a}}n, Rold{\'{a}}n, Moreno{-}Delgado,
  and {\'{A}}lvarez}{Dur{\'{a}}n et~al\mbox{.}}{2014}]{DRMA14}
{\sc Dur{\'{a}}n, F.}, {\sc Rold{\'{a}}n, M.}, {\sc Moreno{-}Delgado, A.}, {\sc
  and} {\sc {\'{A}}lvarez, J.~M.} 2014.
\newblock {Dynamic Validation of Maude Prototypes of UML Models}.
\newblock In {\em {Specification, Algebra, and Software - Essays Dedicated to
  Kokichi Futatsugi (SAS 2014)}}. Lecture Notes in Computer Science, vol. 8373.
  {Springer-Verlag}, 212--228.

\bibitem[\protect\citeauthoryear{Field and Tip}{Field and Tip}{1994}]{FT94}
{\sc Field, J.} {\sc and} {\sc Tip, F.} 1994.
\newblock {Dynamic Dependence in Term rewriting Systems and its Application to
  Program Slicing}.
\newblock In {\em Proc. of the 6th Int'l Symp. on Programming Language
  Implementation and Logic Programming (PLILP 1994)}. Lecture Notes in Computer
  Science, vol. 844. {Springer-Verlag}, 415--431.

\bibitem[\protect\citeauthoryear{Logozzo and Ball}{Logozzo and
  Ball}{2012}]{LB12}
{\sc Logozzo, F.} {\sc and} {\sc Ball, T.} 2012.
\newblock {Modular and Verified Automatic Program Repair}.
\newblock In {\em Proc. of the 27th Annual ACM SIGPLAN Conference on
  Object-Oriented Programming, Systems, Languages, and Applications (OOPSLA
  2012)}. {Association for Computing Machinery}, 133--146.

\bibitem[\protect\citeauthoryear{Mera, L{\'{o}}pez{-}Garc{\'{\i}}a, and
  Hermenegildo}{Mera et~al\mbox{.}}{2009}]{MLH09}
{\sc Mera, E.}, {\sc L{\'{o}}pez{-}Garc{\'{\i}}a, P.}, {\sc and} {\sc
  Hermenegildo, M.~V.} 2009.
\newblock Integrating software testing and run-time checking in an assertion
  verification framework.
\newblock In {\em {Proc. of the 25th Int'l Conference on Logic Programming
  (ICLP 2009)}}. Lecture Notes in Computer Science, vol. 5649.
  {Springer-Verlag}, 281--295.

\bibitem[\protect\citeauthoryear{Meseguer}{Meseguer}{1992}]{Meseguer92}
{\sc Meseguer, J.} 1992.
\newblock {Conditional Rewriting Logic as a Unified Model of Concurrency}.
\newblock {\em Theoretical Computer Science\/}~{\em 96,\/}~1, 73--155.

\bibitem[\protect\citeauthoryear{Ro\c{s}u}{Ro\c{s}u}{2015}]{Rosu15}
{\sc Ro\c{s}u, G.} 2015.
\newblock {From Rewriting Logic, to Programming Language Semantics, to Program
  Verification}.
\newblock In {\em {Logic, Rewriting, and Concurrency - Festschrift Symposium in
  Honor of Jos\'e Meseguer}}. Lecture Notes in Computer Science, vol. 9200.
  {Springer-Verlag}, 598--616.

\end{thebibliography}
\end{document}